\documentclass[apl, aip, reprint]{revtex4-1}
\usepackage{CJK}
\usepackage{graphicx}
\usepackage{amsmath}
\usepackage{amssymb}
\usepackage{color}
\usepackage[english]{babel}
\begin{document}
\begin {CJK*} {GB}{}
\title{Probing Andreev reflection reach \\in semiconductor-superconductor hybrids by Aharonov-Bohm effect}

\author{M. P. Nowak}
\affiliation{AGH University of Science and Technology, Academic Centre for Materials and Nanotechnology, al. A. Mickiewicza 30, 30-059 Krakow, Poland}

\author{P. W{\'o}jcik}
\affiliation{AGH University of Science and Technology, Faculty of Physics and Applied Computer Science, al. A. Mickiewicza 30, 30-059 Krakow, Poland}

\date{\today}

\begin{abstract}
Recent development in fabrication of hybrid nanostructures allows for creation of quantum interferometers that combine semiconductor and superconductor materials. We show that in those nanostructures the joint phenomena of 
Aharonov-Bohm effect and Andreev reflections can be used to determine the length on which the electron is retro-reflected as a hole. We propose to 
exploit this feature for probing of the quasiparticle coherence length in semiconductor-superconductor hybrids by a magnetoconductance measurement.
\end{abstract}

\maketitle
\end{CJK*}
Recently there is a great interest in semiconductor-superconductor hybrids that under proper tuning of the external magnetic field can realize 
topological superconducting phase \cite{sau_generic_2010, oreg_helical_2010} that hosts quasi-particles equivalent of Majorana fermions 
\cite{majorana_symmetric_2006, lutchyn_majorana_2018}. In those hybrids, experimentally realized in the form of quasi 
one-dimensional \cite{chang_hard_2015, krogstrup_epitaxy_2015, gul_hard_2017} or two-dimensional nanowires
\cite{shabani_two-dimensional_2016, kjaergaard_transparent_2017} contacted with a thin superconducting shell, the coupling between electron-like and 
hole-like quasi-particles is induced by the proximity effect microscopically governed by the Andreev reflections. Experimental studies of those 
devices rely on electronic transport measurements where the charge carriers are transfered between the normal and proximitized part. The electrons 
propagating from the semiconducting part with Fermi velocity $v_f$ undergo Andreev reflection at the normal-superconducting (NS) interface on a 
distance corresponding to wave function decay length. In the proximitized part $\Psi \sim\exp[-x/\xi]$, where
\begin{equation}
\xi = \frac{\hbar v_f}{\Delta},
\label{xieq}
\end{equation}
mimics the coherence length of quasi-particles penetrating into the ordinary superconductor.
The reach of Andreev reflection $\xi$ is interesting not only for fundamental reasons, but also it is crucial for the analysis of transport 
measurements as it determines the extent of the structure accessible for probing by the means of tunneling spectroscopy. 
Specifically, recent experiments reported measurement of conductance quantization in Majorana nanowires \cite{mourik_signatures_2012, 
deng_anomalous_2012, finck_anomalous_2013, churchill_superconductor-nanowire_2013, chen_experimental_2017} that signifies ballistic transport over the 
coherence length \cite{kjaergaard_quantized_2016, zhang_ballistic_2017}.

In this letter we propose a unique method to determine the quasiparticle coherence length $\xi$ of the proximitized semiconductor. We exploit 
recent developments in bottom-up synthesis that allows for fabrication of crossing hybrid nanowires \cite{gazibegovic_epitaxy_2017}. 
Nanowire branches can be formed into closed loops creating quantum phase-coherent interferometers with predefined number of epitaxial 
superconductor-semiconductor interfaces. Utilizing combined Aharonov-Bohm \cite{aharonov_significance_1959} effect and Andreev reflection in the topologically trivial phase we exploit this hybrid structures to indirectly visualize the length on which 
the Andreev reflection takes place.

\begin{figure}[ht!]
\center
\includegraphics[width = 7.5cm]{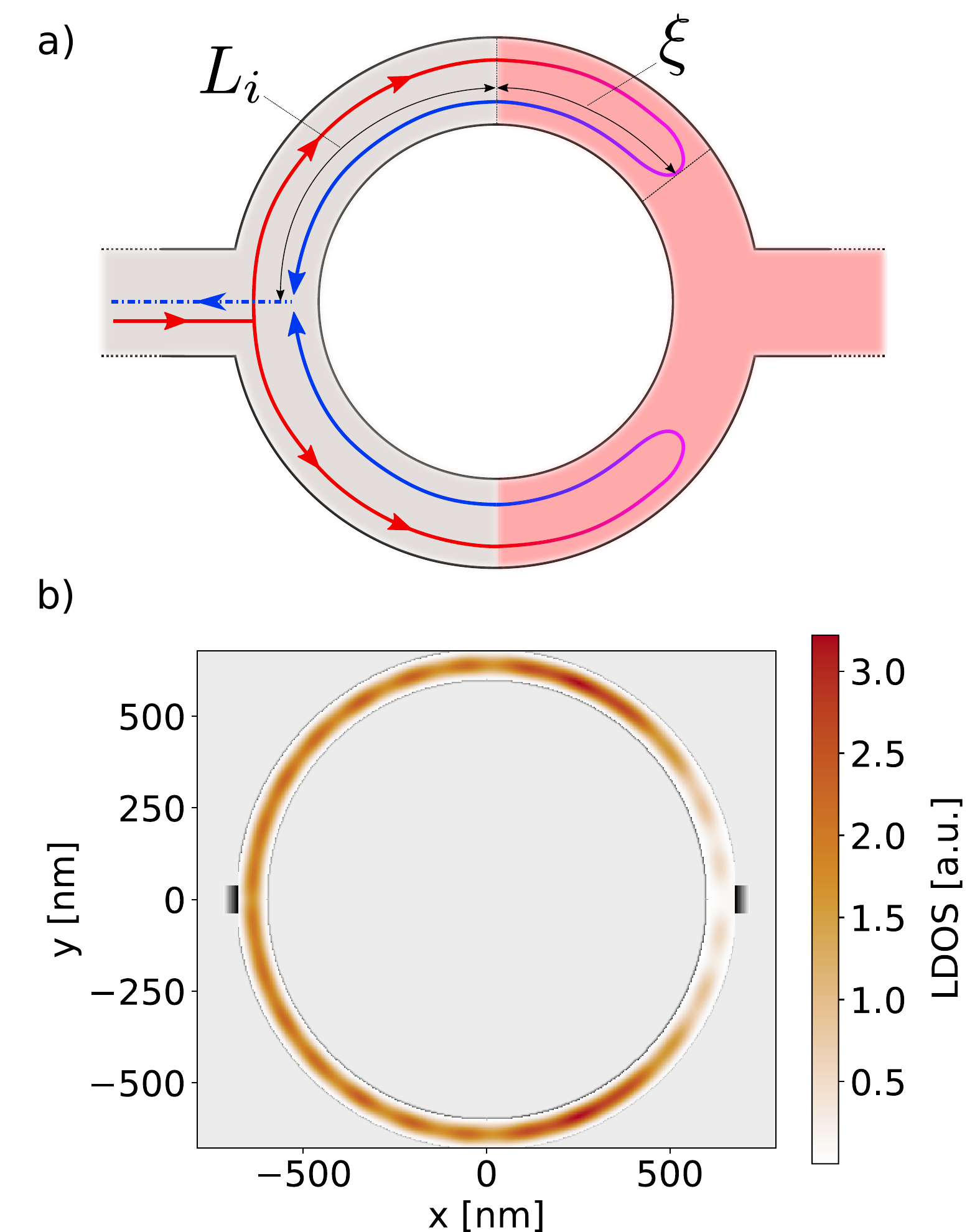}
\caption{a) Cartoon of the quantum interferometer in a form of semiconductor ring proximitized by a superconductor with two semi-infinite leads. The gray and 
pink regions correspond to normal and proximitized parts, respectively. The arrows show propagating electron and hole trajectories with the region where the 
quasiparticle wave function decays in the superconducting part over the coherence length $\xi$. b) Numerically obtained local density of states of 
proximitized interferometer for $\Delta = 0.25$ meV and $B = 5$ mT.}
\label{rings}
\end{figure}

The proposed concept can be explained on an example of quantum ring proximitized by a superconductor [Fig.~\ref{rings}(a)]. In the presence of the magnetic field the transport properties of an ordinary quantum ring is mainly determined by 
the interference effect as the phases $\varphi = \frac{e}{\hbar}\int\mathbf{A}dl$ acquired from the magnetic field by the charged particles 
traveling through the upper and lower arm differ from each other. Consequently, the conductance of the structure undergoes Aharonov-Bohm oscillations 
with the period determined by the flux quanta $\phi_0 = h/e$, as reported experimentally for metallic \cite{webb_observation_1985} and semiconducting \cite{timp_observation_1987} 
quantum rings already in late 80's. Only recently the observation of Aharonov-Bohm effect has been reported for hybrid nanowire networks 
\cite{gazibegovic_epitaxy_2017} which allows for combination of this effect with the scattering processes that occur at the NS interface. 

Let us consider proximitized nanostructure depicted schematically in Fig. \ref{rings} (a). When the excitation energy of the incoming 
electron lies inside the superconducting gap the electron undergoes Andreev reflection at the NS interface. As a result, the period of the Aharonov-Bohm 
oscillations is determined by the phase accumulated both by the propagating electron and the retro-reflected hole. The particles 
acquire phase that correspond to their wave-function span on the total length $L_i^s = 2L_i + 2\xi$, where $2L_i$ is twice the distance of propagation through the normal part of the 
interferometer -- one for scattering of electron, one for the hole. Most importantly the length $L_i^s$ is increased by $2\xi$ that stems
from the phase accumulated by the electron and hole evanescent modes in the proximitized part. The elongation of the effective 
propagation length of the particles by Andreev reflection will lead to a decrease of the Aharonov-Bohm oscillation period. 

\begin{figure}[ht!]
\center
\includegraphics[width = 8.7 cm]{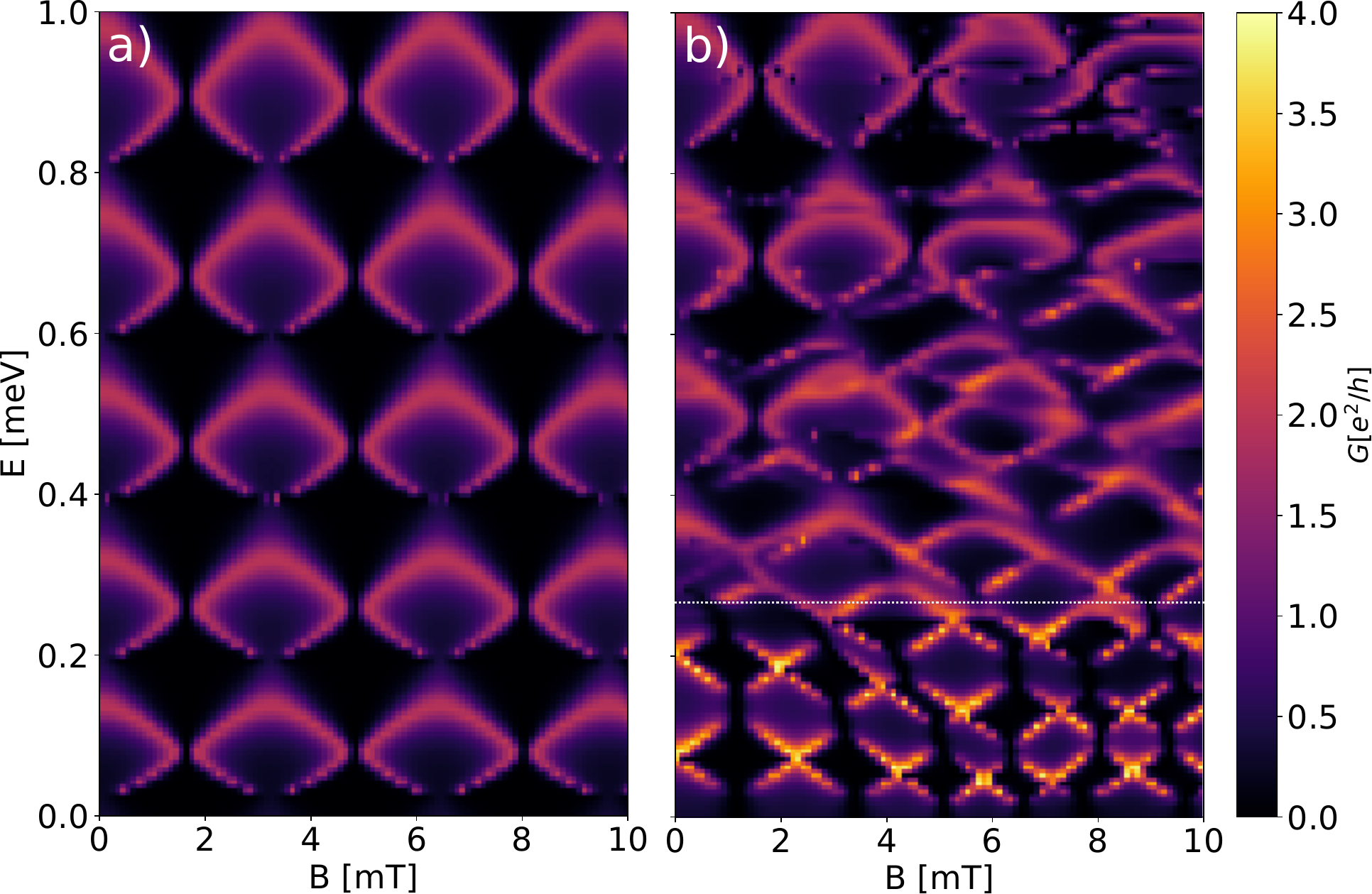}
\caption{Conductance map of a quantum ring versus the excitation energy $E$ of incoming electron and the magnetic field $B$ calculated for 
$\Delta =0$ (a) and $\Delta = 0.25$ meV (b). The dashed horizontal line in (b) denotes the energy of the superconducting gap.}
\label{conductance_maps}
\end{figure}

To test our prediction we perform numerical experiments in which we simulate electronic transport in proximitized interferometers  of different 
geometry in the topologically trivial regime. The system is described by the Bogoliubov-de Gennes Hamiltonian
\begin{equation}
\begin{split}
H = \left( \hbar^2\mathbf{k}^2/2m^*-\mu\right)\sigma_0\tau_z+\Delta\sigma_0\tau_x 
+\alpha(\sigma_xk_y-\sigma_yk_x)\tau_z
\end{split}
\label{2DHamiltonian}
\end{equation}
acting on the spinor $\Psi = (\psi^e_\uparrow, \psi^e_\downarrow, \psi^h_\downarrow, -\psi^h_\uparrow)^\mathrm{T}$, where $e$ ($h$) corresponds to electron-like 
(hole-like) component and $\uparrow$ ($\downarrow$) denotes spin up (down) component. $\Delta$ is the effective induced pairing potential, 
$\sigma_i$ and $\tau_i$ with $i=x,y,z$ are the Pauli matrices acting on spin- and electron-hole degrees of freedom, respectively. The orbital effects 
of the magnetic field are included through the canonical momentum, $\mathbf{k}=-i\nabla+e\mathbf{A}/\hbar\cdot\tau_z$ with the vector-potential in 
the Lorentz gauge $\mathbf{A}=[-yB,0,0]$. As the considered magnetic fields are of order of mT we neglect Zeeman effect and include spin-orbit coupling whose strength is controlled by the parameter $\alpha$. The adopted material parameters correspond to 
recently studied InSb nanostructures \cite{gazibegovic_epitaxy_2017} with $m^*=0.014m$. The numerical problem is solved by discretizing Eq. 
(\ref{2DHamiltonian}) on a square grid with grid spacing $\delta x = \delta y = 4$ nm using Peierls substitution of the hopping elements $t_{nm} 
\rightarrow t_{nm}\exp\left[-ie\int\mathbf{A}d\mathbf{l}/\hbar\right]$ to account for the orbital effects of the magnetic field. We assume that the interferometer is 
connected to semi-infinite leads and calculate the scattering matrix using Kwant \cite{groth_kwant:_2014} package that implements wave function matching method. Finally, we obtain the conductance 
in the linear response regime at zero temperature as $G = e^2/h\cdot(N- R_{ee}+R_{he})$, where $N$ is the number of transverse modes in the leads and 
$R_{ee}$ ($R_{he}$) correspond to the backscattering probability of electrons into electrons (holes).

We start by considering the semiconducting ring with the channel width $W=80$ nm and the radius $R=640$ nm depicted in Fig. \ref{rings} (b). For 
simplicity we neglect the spin-orbit coupling by setting $\alpha = 0$. Firstly we study the case of pure semiconducting ring and assume the 
chemical potential $\mu = 5$ meV such there is one occupied spin-degenerate conducting mode. The map in Fig. \ref{conductance_maps} (a) 
presents conductance as a function of the magnetic field and the incoming electron energy. We observe that the conductance oscillates in $B$ 
independently on the electron energy, with the period $B_p \simeq 3.2$ mT which corresponds to the flux quanta $\phi = BL^2/\pi$ with $L = \pi R$. 
Now let us consider that the ring is half-covered by the superconductor -- however the symmetric coverage of the ring is not a vital assumption of the 
model as we will show later. We set $\Delta = 0.25$ meV and present the corresponding conductance map in Fig. \ref{conductance_maps} (b). We can 
clearly subdivide the map into two regions. For $E>\Delta$ where the excitation energy $E$ exceeds the superconducting gap we see the 
same pattern in conductance oscillations as in the panel (a) but overlayed with resonances on Andreev bound states that are most pronounced at the proximity 
of the gap edge. For $E<\Delta$ (below the white dashed line) where we expect the elongation of the effective propagation length of the particles by $\xi$ due to the Andreev 
reflection, we observe the conductance oscillation pattern with much less period than the one found for $E>\Delta$. The 
magnitude of the maximal conductance is doubled for $E<\Delta$ due to transfer of $2e$ charge through Andreev reflection process 
\cite{beenakker_quantum_1992, kjaergaard_quantized_2016, zhang_ballistic_2017}. Finally, inspecting the local density of states for $B = 5$ mT and $E = 1$ meV in 
Fig. \ref{rings} (b) we observe decaying probability density in the proximitized part as expected from Eq. (\ref{xieq}).

\begin{figure}[ht!]
\center
\includegraphics[width = 7cm]{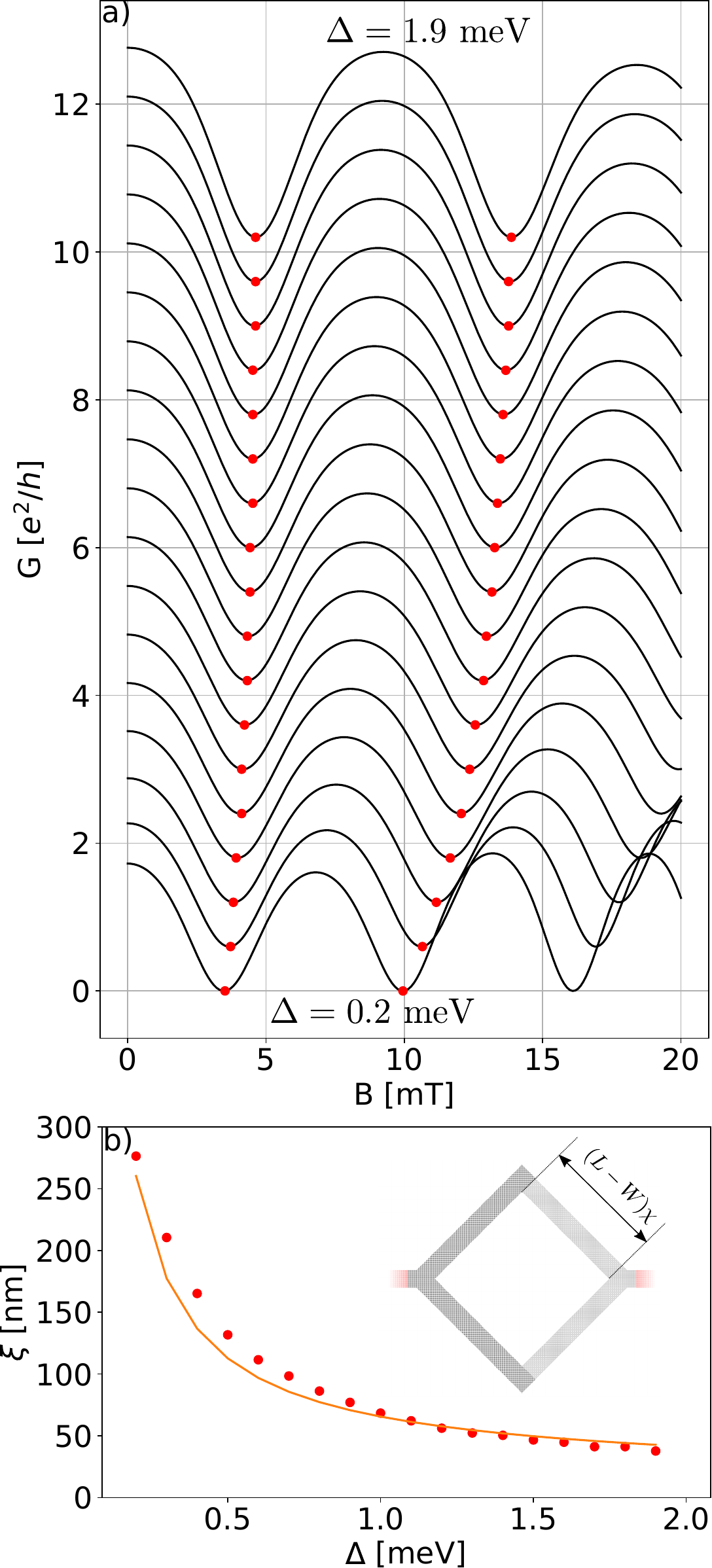}
\caption{a) Conductance traces of a hashtag hybrid interferometer calculated for the superconducting gap varying between $\Delta = 0.2$ meV (bottom curve) and $\Delta = 1.9$ meV (top 
most curve). The traces are sequentially shifted by $0.6$ $e^2/h$ for clarity. The red dots denote the minima of conductance traces used to estimate the Aharonov-Bohm 
oscillation period $B_p$. b) Coherence length estimated from the Aharonov-Bohm oscillation period  according to formula Eq. (\ref{xi_formula}) -- 
red dots -- and the estimates obtained through Eq. (\ref{xieq}) -- orange curve. The inset shows considered interferometer defined in a 
semiconducting hasthag half covered with superconductor (white region), connected with two semi infinite leads.}
\label{comparison}
\end{figure}

Now we turn our attention to the implementation of the discussed concept in state-of-the-art experimental devices. Namely, we consider a 
structure formed by crossing nanowires such they form a hashtag -- a square interferometer \cite{gazibegovic_epitaxy_2017} as presented in the inset 
of Fig.~\ref{comparison} (b). The Aharonov-Bohm conductance oscillations have been already measured in those devices and the development in the 
superconductor deposition allows for an arbitrary arrangement of the superconducting lead -- either during the growth stage 
\cite{krogstrup_epitaxy_2015} or later by the means of litographical deposition \cite{gul_hard_2017}. In our calculations we assume that the device is 
connected to a normal and superconducting leads through two protruding corners of the square and that two arms of the structure are covered by a superconductor on the length $(L-W)\chi$ that opens the energy gap $\Delta$ therein -- see inset in Fig. \ref{comparison} (b). We assume the arm 
length $L=600$ nm, the width $80$ nm and spin-orbit coupling with the strength comparable to the one reported experimentally i.e. $\alpha = 50$ 
meVnm \cite{van_weperen_spin-orbit_2015, kammhuber_conductance_2017, wojcik_tuning_2018}.
 
Figure \ref{comparison} (a) shows conductance traces calculated at $E=0$ for several values of the gap parameter $\Delta$ ranging from $0.2$ meV to $1.9$ meV for the two arms fully covered ($\chi=1$). The traces are sequentially shifted by $0.6$ $e^2/h$ for clarity of the presentation. We see that as the gap is increased the Aharonov-Bohm oscillation period -- the distance between the two red dots -- also increases.

To quantitatively analyze the correspondence between the induced gap parameter $\Delta$ and the Aharonov-Bohm oscillations we compare the coherence length predicted theoretically with the values that can be extracted from the oscillation period. The theoretical estimation of the coherence length is obtained for each transverse mode from Eq. (\ref{xieq}): we consider an infinite channel of width $W$ and obtain the coherence length as the largest decay length $\xi = \max \mathit{Re}[\kappa]^{-1}$ of the evanescent waves $\Psi \sim e^{-\kappa x}$ at zero energy, where $\kappa$ is the eigenvalue of the translational operator \cite{sticlet_robustness_2017, nowak_renormalization_2018}. We plot $\xi$ with the orange curve in Fig. \ref{comparison} (b).

The experimental estimation of the coherence length is cumbersome due to the inability to directly measure the Fermi velocity that depends on the channel geometry, spin-orbit coupling strength, effective mass, etc. Here we demonstrate that one can extract the coherence length form the Aharonov-Bohm oscillation period. By calculating the phase difference acquired by the particles traveling through each arm of the proximitized interferometer \footnote{The choice of a vector potential that minimizes the supercurrent in the considered device is vital for proper description of the orbital effects of the magnetic field, see Ref. \onlinecite{wojcik_durability_2018}} for the considered device geometry we obtain the formula for the coherence length,
\begin{equation}
\xi = (L-W)\chi + W/2 - \sqrt{2L^2 - \frac{2\pi\hbar}{eB_p}}.
\label{xi_formula}
\end{equation}
Taking $\chi=1$ for the device from the inset of Fig. \ref{comparison} (b) we plot extracted coherence length in the panel (b) with the red dots and observe a very good agreement with the theoretically predicted coherence lengths plotted with the orange curve. 

\begin{figure}[ht!]
\center
\includegraphics[width = 7cm]{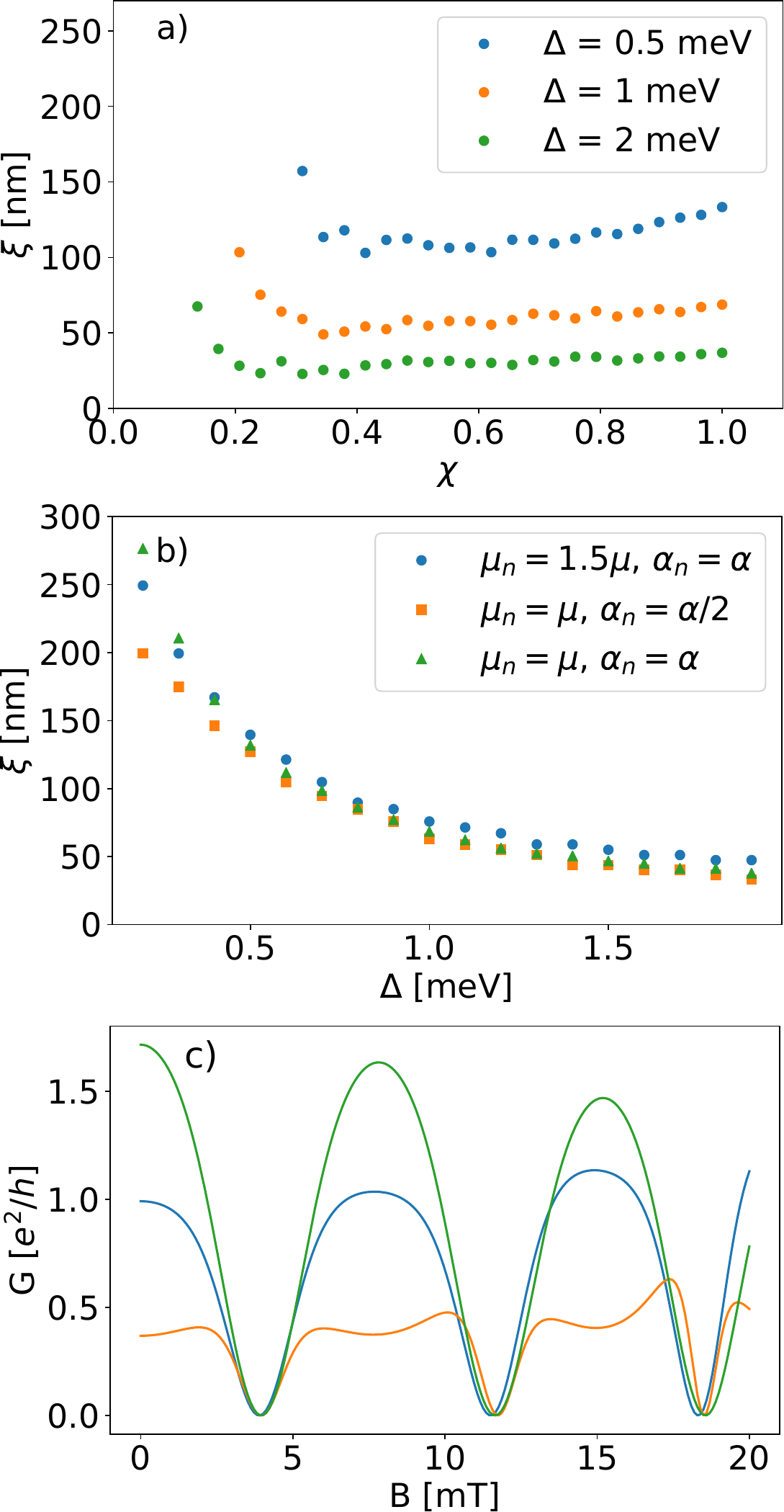}
\caption{a) Estimated coherence length versus the amount of coverage of the interferometer arm by the superconductor $\chi$ for three values of the induced gap. b) Estimated coherence length for a system where effective parameters differ between the normal ($\mu_n, \alpha_n$) and the proximitized regions. c) Conductance versus the magnetic field for the parameters of the normal part from the panel (b).}
\label{varied}
\end{figure}

Furthermore, Eq. (\ref{xi_formula}) allows for extraction of the coherence length for a device with arbitrary amount of the arm coverage. In Fig. \ref{varied} (a) we show $\xi$ obtained for three values of the induced gap parameter when the arm coverage parameter $\chi$ changes. We observe that the obtained $\xi$ values clearly differ for the three values of $\Delta$ and that they remain approximately constant as the coverage parameter is varied. The highest variation is obtained for small $\chi$ when the coherence length becomes comparable with the length of the covered part.

Finally, we test the robustness of the proposed method with respect to the variation of effective parameters along the structure. Coupling to the supercondcuting shell modifies parameters of the covered semiconductor which results in variation of the electronic properties along the structure  \cite{antipov_effects_2018, winkler_unified_2018}. This effect might lead to non-transparent interface between the uncovered and proximitized semiconductor. Note however that already transport experiments done on 2DEG and nanowire NS junctions showed that such an interface is pristine\cite{kjaergaard_transparent_2017, zhang_ballistic_2017}. To account for possible effects of the parameter variation we performed k.p calculations [see Refs. \onlinecite{wojcik_tuning_2018, wojcik_enhanced_2018} for the details] and estimated that Rashba spin-orbit coupling parameter differ between the proximitized and normal regions up to a factor of two. In Fig. \ref{varied} (b) with orange squares we plot the estimated coherence length obtained for twice weaker spin-orbit coupling in the normal part which remain very close to the values obtained for the case of spatially constant parameters plotted with the green triangles. Also when we decrease the chemical potential in the normal part the extracted coherence length remains to a large degree unchanged -- see the blue circles in Fig. \ref{varied} (b). In fact, varied parameters at the interface decrease Andreev reflection probability and by that the conductance -- see the decreased magnitude of the blue and orange conductance traces in Fig. \ref{varied} (c). This process however does not modify the Aharonov-Bohm oscillation period that reflects the coherence length.

In summary we considered the effect of Andreev reflection in the quantum interferometer partially proximitized by superconductor in the presence of the external magnetic field. We showed that the period of the Aharonov-Bohm conductance oscillations depends on the electron and hole 
wavefunctions decay length $\xi$ in the proximitized part allowing to probe the length on which the Andreev reflection occur in experimentally realizable devices. We proposed to exploit this feature for probing of quasiparticle coherence length by magneto-conductance measurement and confronted this idea with numerical quantum transport simulations of quantum rings and nanowire hashtags. The method that we propose here bases on easily achievable conductance measurement of the Aharonov-Bohm oscillations which renders it as a feasible mean for experimentally probing the quasiparticle coherence length in proximitized nanostructures.

This work was supported by National Science Centre (NCN) decision number DEC-2016/23/D/ST3/00394 and partially by Academic Centre for Materials and 
Nanotechnology AGH UST statutory tasks within subsidy of Ministry of Science and Higher Education. The calculations were performed on PL-Grid 
Infrastructure.


\begin{thebibliography}{32}%
\makeatletter
\providecommand \@ifxundefined [1]{%
 \@ifx{#1\undefined}
}%
\providecommand \@ifnum [1]{%
 \ifnum #1\expandafter \@firstoftwo
 \else \expandafter \@secondoftwo
 \fi
}%
\providecommand \@ifx [1]{%
 \ifx #1\expandafter \@firstoftwo
 \else \expandafter \@secondoftwo
 \fi
}%
\providecommand \natexlab [1]{#1}%
\providecommand \enquote  [1]{``#1''}%
\providecommand \bibnamefont  [1]{#1}%
\providecommand \bibfnamefont [1]{#1}%
\providecommand \citenamefont [1]{#1}%
\providecommand \href@noop [0]{\@secondoftwo}%
\providecommand \href [0]{\begingroup \@sanitize@url \@href}%
\providecommand \@href[1]{\@@startlink{#1}\@@href}%
\providecommand \@@href[1]{\endgroup#1\@@endlink}%
\providecommand \@sanitize@url [0]{\catcode `\\12\catcode `\$12\catcode
  `\&12\catcode `\#12\catcode `\^12\catcode `\_12\catcode `\%12\relax}%
\providecommand \@@startlink[1]{}%
\providecommand \@@endlink[0]{}%
\providecommand \url  [0]{\begingroup\@sanitize@url \@url }%
\providecommand \@url [1]{\endgroup\@href {#1}{\urlprefix }}%
\providecommand \urlprefix  [0]{URL }%
\providecommand \Eprint [0]{\href }%
\providecommand \doibase [0]{http://dx.doi.org/}%
\providecommand \selectlanguage [0]{\@gobble}%
\providecommand \bibinfo  [0]{\@secondoftwo}%
\providecommand \bibfield  [0]{\@secondoftwo}%
\providecommand \translation [1]{[#1]}%
\providecommand \BibitemOpen [0]{}%
\providecommand \bibitemStop [0]{}%
\providecommand \bibitemNoStop [0]{.\EOS\space}%
\providecommand \EOS [0]{\spacefactor3000\relax}%
\providecommand \BibitemShut  [1]{\csname bibitem#1\endcsname}%
\let\auto@bib@innerbib\@empty
\bibitem [{\citenamefont {Sau}\ \emph {et~al.}(2010)\citenamefont {Sau},
  \citenamefont {Lutchyn}, \citenamefont {Tewari},\ and\ \citenamefont
  {Das~Sarma}}]{sau_generic_2010}%
  \BibitemOpen
  \bibfield  {author} {\bibinfo {author} {\bibfnamefont {J.~D.}\ \bibnamefont
  {Sau}}, \bibinfo {author} {\bibfnamefont {R.~M.}\ \bibnamefont {Lutchyn}},
  \bibinfo {author} {\bibfnamefont {S.}~\bibnamefont {Tewari}}, \ and\ \bibinfo
  {author} {\bibfnamefont {S.}~\bibnamefont {Das~Sarma}},\ }\href {\doibase
  10.1103/PhysRevLett.104.040502} {\bibfield  {journal} {\bibinfo  {journal}
  {Phys. Rev. Lett.}\ }\textbf {\bibinfo {volume} {104}},\ \bibinfo {pages}
  {040502} (\bibinfo {year} {2010})}\BibitemShut {NoStop}%
\bibitem [{\citenamefont {Oreg}, \citenamefont {Refael},\ and\ \citenamefont
  {von Oppen}(2010)}]{oreg_helical_2010}%
  \BibitemOpen
  \bibfield  {author} {\bibinfo {author} {\bibfnamefont {Y.}~\bibnamefont
  {Oreg}}, \bibinfo {author} {\bibfnamefont {G.}~\bibnamefont {Refael}}, \ and\
  \bibinfo {author} {\bibfnamefont {F.}~\bibnamefont {von Oppen}},\ }\href
  {\doibase 10.1103/PhysRevLett.105.177002} {\bibfield  {journal} {\bibinfo
  {journal} {Phys. Rev. Lett.}\ }\textbf {\bibinfo {volume} {105}},\ \bibinfo
  {pages} {177002} (\bibinfo {year} {2010})}\BibitemShut {NoStop}%
\bibitem [{\citenamefont {Majorana}\ and\ \citenamefont
  {Maiani}(2006)}]{majorana_symmetric_2006}%
  \BibitemOpen
  \bibfield  {author} {\bibinfo {author} {\bibfnamefont {E.}~\bibnamefont
  {Majorana}}\ and\ \bibinfo {author} {\bibfnamefont {L.}~\bibnamefont
  {Maiani}},\ }in\ \href {\doibase 10.1007/978-3-540-48095-2_10} {\emph
  {\bibinfo {booktitle} {Ettore {Majorana} {Scientific} {Papers}}}}\ (\bibinfo
  {publisher} {Springer, Berlin, Heidelberg},\ \bibinfo {year} {2006})\ pp.\
  \bibinfo {pages} {201--233}\BibitemShut {NoStop}%
\bibitem [{\citenamefont {Lutchyn}\ \emph {et~al.}(2018)\citenamefont
  {Lutchyn}, \citenamefont {Bakkers}, \citenamefont {Kouwenhoven},
  \citenamefont {Krogstrup}, \citenamefont {Marcus},\ and\ \citenamefont
  {Oreg}}]{lutchyn_majorana_2018}%
  \BibitemOpen
  \bibfield  {author} {\bibinfo {author} {\bibfnamefont {R.~M.}\ \bibnamefont
  {Lutchyn}}, \bibinfo {author} {\bibfnamefont {E.~P. a.~M.}\ \bibnamefont
  {Bakkers}}, \bibinfo {author} {\bibfnamefont {L.~P.}\ \bibnamefont
  {Kouwenhoven}}, \bibinfo {author} {\bibfnamefont {P.}~\bibnamefont
  {Krogstrup}}, \bibinfo {author} {\bibfnamefont {C.~M.}\ \bibnamefont
  {Marcus}}, \ and\ \bibinfo {author} {\bibfnamefont {Y.}~\bibnamefont
  {Oreg}},\ }\href {\doibase 10.1038/s41578-018-0003-1} {\bibfield  {journal}
  {\bibinfo  {journal} {Nature Reviews Materials}\ }\textbf {\bibinfo {volume}
  {3}},\ \bibinfo {pages} {52} (\bibinfo {year} {2018})}\BibitemShut {NoStop}%
\bibitem [{\citenamefont {Chang}\ \emph {et~al.}(2015)\citenamefont {Chang},
  \citenamefont {Albrecht}, \citenamefont {Jespersen}, \citenamefont
  {Kuemmeth}, \citenamefont {Krogstrup}, \citenamefont {Nyg{\r a}rd},\ and\
  \citenamefont {Marcus}}]{chang_hard_2015}%
  \BibitemOpen
  \bibfield  {author} {\bibinfo {author} {\bibfnamefont {W.}~\bibnamefont
  {Chang}}, \bibinfo {author} {\bibfnamefont {S.~M.}\ \bibnamefont {Albrecht}},
  \bibinfo {author} {\bibfnamefont {T.~S.}\ \bibnamefont {Jespersen}}, \bibinfo
  {author} {\bibfnamefont {F.}~\bibnamefont {Kuemmeth}}, \bibinfo {author}
  {\bibfnamefont {P.}~\bibnamefont {Krogstrup}}, \bibinfo {author}
  {\bibfnamefont {J.}~\bibnamefont {Nyg{\r a}rd}}, \ and\ \bibinfo {author}
  {\bibfnamefont {C.~M.}\ \bibnamefont {Marcus}},\ }\href {\doibase
  10.1038/nnano.2014.306} {\bibfield  {journal} {\bibinfo  {journal} {Nature
  Nano.}\ }\textbf {\bibinfo {volume} {10}},\ \bibinfo {pages} {232} (\bibinfo
  {year} {2015})}\BibitemShut {NoStop}%
\bibitem [{\citenamefont {Krogstrup}\ \emph {et~al.}(2015)\citenamefont
  {Krogstrup}, \citenamefont {Ziino}, \citenamefont {Chang}, \citenamefont
  {Albrecht}, \citenamefont {Madsen}, \citenamefont {Johnson}, \citenamefont
  {Nyg{\r a}rd}, \citenamefont {Marcus},\ and\ \citenamefont
  {Jespersen}}]{krogstrup_epitaxy_2015}%
  \BibitemOpen
  \bibfield  {author} {\bibinfo {author} {\bibfnamefont {P.}~\bibnamefont
  {Krogstrup}}, \bibinfo {author} {\bibfnamefont {N.~L.~B.}\ \bibnamefont
  {Ziino}}, \bibinfo {author} {\bibfnamefont {W.}~\bibnamefont {Chang}},
  \bibinfo {author} {\bibfnamefont {S.~M.}\ \bibnamefont {Albrecht}}, \bibinfo
  {author} {\bibfnamefont {M.~H.}\ \bibnamefont {Madsen}}, \bibinfo {author}
  {\bibfnamefont {E.}~\bibnamefont {Johnson}}, \bibinfo {author} {\bibfnamefont
  {J.}~\bibnamefont {Nyg{\r a}rd}}, \bibinfo {author} {\bibfnamefont {C.~M.}\
  \bibnamefont {Marcus}}, \ and\ \bibinfo {author} {\bibfnamefont {T.~S.}\
  \bibnamefont {Jespersen}},\ }\href {\doibase 10.1038/nmat4176} {\bibfield
  {journal} {\bibinfo  {journal} {Nature Materials}\ }\textbf {\bibinfo
  {volume} {14}},\ \bibinfo {pages} {400} (\bibinfo {year} {2015})}\BibitemShut
  {NoStop}%
\bibitem [{\citenamefont {G{\"u}l}\ \emph {et~al.}(2017)\citenamefont
  {G{\"u}l}, \citenamefont {Zhang}, \citenamefont {de~Vries}, \citenamefont
  {van Veen}, \citenamefont {Zuo}, \citenamefont {Mourik}, \citenamefont
  {Conesa-Boj}, \citenamefont {Nowak}, \citenamefont {van Woerkom},
  \citenamefont {Quintero-P{\'e}rez}, \citenamefont {Cassidy}, \citenamefont
  {Geresdi}, \citenamefont {Koelling}, \citenamefont {Car}, \citenamefont
  {Plissard}, \citenamefont {Bakkers},\ and\ \citenamefont
  {Kouwenhoven}}]{gul_hard_2017}%
  \BibitemOpen
  \bibfield  {author} {\bibinfo {author} {\bibfnamefont {{\"O}.}~\bibnamefont
  {G{\"u}l}}, \bibinfo {author} {\bibfnamefont {H.}~\bibnamefont {Zhang}},
  \bibinfo {author} {\bibfnamefont {F.~K.}\ \bibnamefont {de~Vries}}, \bibinfo
  {author} {\bibfnamefont {J.}~\bibnamefont {van Veen}}, \bibinfo {author}
  {\bibfnamefont {K.}~\bibnamefont {Zuo}}, \bibinfo {author} {\bibfnamefont
  {V.}~\bibnamefont {Mourik}}, \bibinfo {author} {\bibfnamefont
  {S.}~\bibnamefont {Conesa-Boj}}, \bibinfo {author} {\bibfnamefont {M.~P.}\
  \bibnamefont {Nowak}}, \bibinfo {author} {\bibfnamefont {D.~J.}\ \bibnamefont
  {van Woerkom}}, \bibinfo {author} {\bibfnamefont {M.}~\bibnamefont
  {Quintero-P{\'e}rez}}, \bibinfo {author} {\bibfnamefont {M.~C.}\ \bibnamefont
  {Cassidy}}, \bibinfo {author} {\bibfnamefont {A.}~\bibnamefont {Geresdi}},
  \bibinfo {author} {\bibfnamefont {S.}~\bibnamefont {Koelling}}, \bibinfo
  {author} {\bibfnamefont {D.}~\bibnamefont {Car}}, \bibinfo {author}
  {\bibfnamefont {S.~R.}\ \bibnamefont {Plissard}}, \bibinfo {author}
  {\bibfnamefont {E.~P. A.~M.}\ \bibnamefont {Bakkers}}, \ and\ \bibinfo
  {author} {\bibfnamefont {L.~P.}\ \bibnamefont {Kouwenhoven}},\ }\href
  {\doibase 10.1021/acs.nanolett.7b00540} {\bibfield  {journal} {\bibinfo
  {journal} {Nano Lett.}\ }\textbf {\bibinfo {volume} {17}},\ \bibinfo {pages}
  {2690} (\bibinfo {year} {2017})}\BibitemShut {NoStop}%
\bibitem [{\citenamefont {Shabani}\ \emph {et~al.}(2016)\citenamefont
  {Shabani}, \citenamefont {Kjaergaard}, \citenamefont {Suominen},
  \citenamefont {Kim}, \citenamefont {Nichele}, \citenamefont {Pakrouski},
  \citenamefont {Stankevic}, \citenamefont {Lutchyn}, \citenamefont
  {Krogstrup}, \citenamefont {Feidenhans'l}, \citenamefont {Kraemer},
  \citenamefont {Nayak}, \citenamefont {Troyer}, \citenamefont {Marcus},\ and\
  \citenamefont {Palmstr{\o}m}}]{shabani_two-dimensional_2016}%
  \BibitemOpen
  \bibfield  {author} {\bibinfo {author} {\bibfnamefont {J.}~\bibnamefont
  {Shabani}}, \bibinfo {author} {\bibfnamefont {M.}~\bibnamefont {Kjaergaard}},
  \bibinfo {author} {\bibfnamefont {H.~J.}\ \bibnamefont {Suominen}}, \bibinfo
  {author} {\bibfnamefont {Y.}~\bibnamefont {Kim}}, \bibinfo {author}
  {\bibfnamefont {F.}~\bibnamefont {Nichele}}, \bibinfo {author} {\bibfnamefont
  {K.}~\bibnamefont {Pakrouski}}, \bibinfo {author} {\bibfnamefont
  {T.}~\bibnamefont {Stankevic}}, \bibinfo {author} {\bibfnamefont {R.~M.}\
  \bibnamefont {Lutchyn}}, \bibinfo {author} {\bibfnamefont {P.}~\bibnamefont
  {Krogstrup}}, \bibinfo {author} {\bibfnamefont {R.}~\bibnamefont
  {Feidenhans'l}}, \bibinfo {author} {\bibfnamefont {S.}~\bibnamefont
  {Kraemer}}, \bibinfo {author} {\bibfnamefont {C.}~\bibnamefont {Nayak}},
  \bibinfo {author} {\bibfnamefont {M.}~\bibnamefont {Troyer}}, \bibinfo
  {author} {\bibfnamefont {C.~M.}\ \bibnamefont {Marcus}}, \ and\ \bibinfo
  {author} {\bibfnamefont {C.~J.}\ \bibnamefont {Palmstr{\o}m}},\ }\href
  {\doibase 10.1103/PhysRevB.93.155402} {\bibfield  {journal} {\bibinfo
  {journal} {Phys. Rev. B}\ }\textbf {\bibinfo {volume} {93}},\ \bibinfo
  {pages} {155402} (\bibinfo {year} {2016})}\BibitemShut {NoStop}%
\bibitem [{\citenamefont {Kjaergaard}\ \emph {et~al.}(2017)\citenamefont
  {Kjaergaard}, \citenamefont {Suominen}, \citenamefont {Nowak}, \citenamefont
  {Akhmerov}, \citenamefont {Shabani}, \citenamefont {Palmstr{\o}m},
  \citenamefont {Nichele},\ and\ \citenamefont
  {Marcus}}]{kjaergaard_transparent_2017}%
  \BibitemOpen
  \bibfield  {author} {\bibinfo {author} {\bibfnamefont {M.}~\bibnamefont
  {Kjaergaard}}, \bibinfo {author} {\bibfnamefont {H.~J.}\ \bibnamefont
  {Suominen}}, \bibinfo {author} {\bibfnamefont {M.~P.}\ \bibnamefont {Nowak}},
  \bibinfo {author} {\bibfnamefont {A.~R.}\ \bibnamefont {Akhmerov}}, \bibinfo
  {author} {\bibfnamefont {J.}~\bibnamefont {Shabani}}, \bibinfo {author}
  {\bibfnamefont {C.~J.}\ \bibnamefont {Palmstr{\o}m}}, \bibinfo {author}
  {\bibfnamefont {F.}~\bibnamefont {Nichele}}, \ and\ \bibinfo {author}
  {\bibfnamefont {C.~M.}\ \bibnamefont {Marcus}},\ }\href {\doibase
  10.1103/PhysRevApplied.7.034029} {\bibfield  {journal} {\bibinfo  {journal}
  {Phys. Rev. Applied}\ }\textbf {\bibinfo {volume} {7}},\ \bibinfo {pages}
  {034029} (\bibinfo {year} {2017})}\BibitemShut {NoStop}%
\bibitem [{\citenamefont {Mourik}\ \emph {et~al.}(2012)\citenamefont {Mourik},
  \citenamefont {Zuo}, \citenamefont {Frolov}, \citenamefont {Plissard},
  \citenamefont {Bakkers},\ and\ \citenamefont
  {Kouwenhoven}}]{mourik_signatures_2012}%
  \BibitemOpen
  \bibfield  {author} {\bibinfo {author} {\bibfnamefont {V.}~\bibnamefont
  {Mourik}}, \bibinfo {author} {\bibfnamefont {K.}~\bibnamefont {Zuo}},
  \bibinfo {author} {\bibfnamefont {S.~M.}\ \bibnamefont {Frolov}}, \bibinfo
  {author} {\bibfnamefont {S.~R.}\ \bibnamefont {Plissard}}, \bibinfo {author}
  {\bibfnamefont {E.~P. a.~M.}\ \bibnamefont {Bakkers}}, \ and\ \bibinfo
  {author} {\bibfnamefont {L.~P.}\ \bibnamefont {Kouwenhoven}},\ }\href
  {\doibase 10.1126/science.1222360} {\bibfield  {journal} {\bibinfo  {journal}
  {Science}\ }\textbf {\bibinfo {volume} {336}},\ \bibinfo {pages} {1003}
  (\bibinfo {year} {2012})}\BibitemShut {NoStop}%
\bibitem [{\citenamefont {Deng}\ \emph {et~al.}(2012)\citenamefont {Deng},
  \citenamefont {Yu}, \citenamefont {Huang}, \citenamefont {Larsson},
  \citenamefont {Caroff},\ and\ \citenamefont {Xu}}]{deng_anomalous_2012}%
  \BibitemOpen
  \bibfield  {author} {\bibinfo {author} {\bibfnamefont {M.~T.}\ \bibnamefont
  {Deng}}, \bibinfo {author} {\bibfnamefont {C.~L.}\ \bibnamefont {Yu}},
  \bibinfo {author} {\bibfnamefont {G.~Y.}\ \bibnamefont {Huang}}, \bibinfo
  {author} {\bibfnamefont {M.}~\bibnamefont {Larsson}}, \bibinfo {author}
  {\bibfnamefont {P.}~\bibnamefont {Caroff}}, \ and\ \bibinfo {author}
  {\bibfnamefont {H.~Q.}\ \bibnamefont {Xu}},\ }\href {\doibase
  10.1021/nl303758w} {\bibfield  {journal} {\bibinfo  {journal} {Nano Lett.}\
  }\textbf {\bibinfo {volume} {12}},\ \bibinfo {pages} {6414} (\bibinfo {year}
  {2012})}\BibitemShut {NoStop}%
\bibitem [{\citenamefont {Finck}\ \emph {et~al.}(2013)\citenamefont {Finck},
  \citenamefont {Van~Harlingen}, \citenamefont {Mohseni}, \citenamefont
  {Jung},\ and\ \citenamefont {Li}}]{finck_anomalous_2013}%
  \BibitemOpen
  \bibfield  {author} {\bibinfo {author} {\bibfnamefont {A.~D.~K.}\
  \bibnamefont {Finck}}, \bibinfo {author} {\bibfnamefont {D.~J.}\ \bibnamefont
  {Van~Harlingen}}, \bibinfo {author} {\bibfnamefont {P.~K.}\ \bibnamefont
  {Mohseni}}, \bibinfo {author} {\bibfnamefont {K.}~\bibnamefont {Jung}}, \
  and\ \bibinfo {author} {\bibfnamefont {X.}~\bibnamefont {Li}},\ }\href
  {\doibase 10.1103/PhysRevLett.110.126406} {\bibfield  {journal} {\bibinfo
  {journal} {Phys. Rev. Lett.}\ }\textbf {\bibinfo {volume} {110}},\ \bibinfo
  {pages} {126406} (\bibinfo {year} {2013})}\BibitemShut {NoStop}%
\bibitem [{\citenamefont {Churchill}\ \emph {et~al.}(2013)\citenamefont
  {Churchill}, \citenamefont {Fatemi}, \citenamefont {Grove-Rasmussen},
  \citenamefont {Deng}, \citenamefont {Caroff}, \citenamefont {Xu},\ and\
  \citenamefont {Marcus}}]{churchill_superconductor-nanowire_2013}%
  \BibitemOpen
  \bibfield  {author} {\bibinfo {author} {\bibfnamefont {H.~O.~H.}\
  \bibnamefont {Churchill}}, \bibinfo {author} {\bibfnamefont {V.}~\bibnamefont
  {Fatemi}}, \bibinfo {author} {\bibfnamefont {K.}~\bibnamefont
  {Grove-Rasmussen}}, \bibinfo {author} {\bibfnamefont {M.~T.}\ \bibnamefont
  {Deng}}, \bibinfo {author} {\bibfnamefont {P.}~\bibnamefont {Caroff}},
  \bibinfo {author} {\bibfnamefont {H.~Q.}\ \bibnamefont {Xu}}, \ and\ \bibinfo
  {author} {\bibfnamefont {C.~M.}\ \bibnamefont {Marcus}},\ }\href {\doibase
  10.1103/PhysRevB.87.241401} {\bibfield  {journal} {\bibinfo  {journal} {Phys.
  Rev. B}\ }\textbf {\bibinfo {volume} {87}},\ \bibinfo {pages} {241401}
  (\bibinfo {year} {2013})}\BibitemShut {NoStop}%
\bibitem [{\citenamefont {Chen}\ \emph {et~al.}(2017)\citenamefont {Chen},
  \citenamefont {Yu}, \citenamefont {Stenger}, \citenamefont {Hocevar},
  \citenamefont {Car}, \citenamefont {Plissard}, \citenamefont {Bakkers},
  \citenamefont {Stanescu},\ and\ \citenamefont
  {Frolov}}]{chen_experimental_2017}%
  \BibitemOpen
  \bibfield  {author} {\bibinfo {author} {\bibfnamefont {J.}~\bibnamefont
  {Chen}}, \bibinfo {author} {\bibfnamefont {P.}~\bibnamefont {Yu}}, \bibinfo
  {author} {\bibfnamefont {J.}~\bibnamefont {Stenger}}, \bibinfo {author}
  {\bibfnamefont {M.}~\bibnamefont {Hocevar}}, \bibinfo {author} {\bibfnamefont
  {D.}~\bibnamefont {Car}}, \bibinfo {author} {\bibfnamefont {S.~R.}\
  \bibnamefont {Plissard}}, \bibinfo {author} {\bibfnamefont {E.~P. A.~M.}\
  \bibnamefont {Bakkers}}, \bibinfo {author} {\bibfnamefont {T.~D.}\
  \bibnamefont {Stanescu}}, \ and\ \bibinfo {author} {\bibfnamefont {S.~M.}\
  \bibnamefont {Frolov}},\ }\href {\doibase 10.1126/sciadv.1701476} {\bibfield
  {journal} {\bibinfo  {journal} {Science Advances}\ }\textbf {\bibinfo
  {volume} {3}},\ \bibinfo {pages} {e1701476} (\bibinfo {year}
  {2017})}\BibitemShut {NoStop}%
\bibitem [{\citenamefont {Kjaergaard}\ \emph {et~al.}(2016)\citenamefont
  {Kjaergaard}, \citenamefont {Nichele}, \citenamefont {Suominen},
  \citenamefont {Nowak}, \citenamefont {Wimmer}, \citenamefont {Akhmerov},
  \citenamefont {Folk}, \citenamefont {Flensberg}, \citenamefont {Shabani},
  \citenamefont {Palmstr{\o}m},\ and\ \citenamefont
  {Marcus}}]{kjaergaard_quantized_2016}%
  \BibitemOpen
  \bibfield  {author} {\bibinfo {author} {\bibfnamefont {M.}~\bibnamefont
  {Kjaergaard}}, \bibinfo {author} {\bibfnamefont {F.}~\bibnamefont {Nichele}},
  \bibinfo {author} {\bibfnamefont {H.~J.}\ \bibnamefont {Suominen}}, \bibinfo
  {author} {\bibfnamefont {M.~P.}\ \bibnamefont {Nowak}}, \bibinfo {author}
  {\bibfnamefont {M.}~\bibnamefont {Wimmer}}, \bibinfo {author} {\bibfnamefont
  {A.~R.}\ \bibnamefont {Akhmerov}}, \bibinfo {author} {\bibfnamefont {J.~A.}\
  \bibnamefont {Folk}}, \bibinfo {author} {\bibfnamefont {K.}~\bibnamefont
  {Flensberg}}, \bibinfo {author} {\bibfnamefont {J.}~\bibnamefont {Shabani}},
  \bibinfo {author} {\bibfnamefont {C.~J.}\ \bibnamefont {Palmstr{\o}m}}, \
  and\ \bibinfo {author} {\bibfnamefont {C.~M.}\ \bibnamefont {Marcus}},\
  }\href {\doibase 10.1038/ncomms12841} {\bibfield  {journal} {\bibinfo
  {journal} {Nature Commun.}\ }\textbf {\bibinfo {volume} {7}},\ \bibinfo
  {pages} {12841} (\bibinfo {year} {2016})}\BibitemShut {NoStop}%
\bibitem [{\citenamefont {Zhang}\ \emph {et~al.}(2017)\citenamefont {Zhang},
  \citenamefont {G{\"u}l}, \citenamefont {Conesa-Boj}, \citenamefont {Nowak},
  \citenamefont {Wimmer}, \citenamefont {Zuo}, \citenamefont {Mourik},
  \citenamefont {Vries}, \citenamefont {Veen}, \citenamefont {Moor},
  \citenamefont {Bommer}, \citenamefont {Woerkom}, \citenamefont {Car},
  \citenamefont {Plissard}, \citenamefont {Bakkers}, \citenamefont
  {Quintero-P{\'e}rez}, \citenamefont {Cassidy}, \citenamefont {Koelling},
  \citenamefont {Goswami}, \citenamefont {Watanabe}, \citenamefont
  {Taniguchi},\ and\ \citenamefont {Kouwenhoven}}]{zhang_ballistic_2017}%
  \BibitemOpen
  \bibfield  {author} {\bibinfo {author} {\bibfnamefont {H.}~\bibnamefont
  {Zhang}}, \bibinfo {author} {\bibfnamefont {{\"O}.}~\bibnamefont {G{\"u}l}},
  \bibinfo {author} {\bibfnamefont {S.}~\bibnamefont {Conesa-Boj}}, \bibinfo
  {author} {\bibfnamefont {M.~P.}\ \bibnamefont {Nowak}}, \bibinfo {author}
  {\bibfnamefont {M.}~\bibnamefont {Wimmer}}, \bibinfo {author} {\bibfnamefont
  {K.}~\bibnamefont {Zuo}}, \bibinfo {author} {\bibfnamefont {V.}~\bibnamefont
  {Mourik}}, \bibinfo {author} {\bibfnamefont {F.~K.~d.}\ \bibnamefont
  {Vries}}, \bibinfo {author} {\bibfnamefont {J.~v.}\ \bibnamefont {Veen}},
  \bibinfo {author} {\bibfnamefont {M.~W. A.~d.}\ \bibnamefont {Moor}},
  \bibinfo {author} {\bibfnamefont {J.~D.~S.}\ \bibnamefont {Bommer}}, \bibinfo
  {author} {\bibfnamefont {D.~J.~v.}\ \bibnamefont {Woerkom}}, \bibinfo
  {author} {\bibfnamefont {D.}~\bibnamefont {Car}}, \bibinfo {author}
  {\bibfnamefont {S.~R.}\ \bibnamefont {Plissard}}, \bibinfo {author}
  {\bibfnamefont {E.~P. A.~M.}\ \bibnamefont {Bakkers}}, \bibinfo {author}
  {\bibfnamefont {M.}~\bibnamefont {Quintero-P{\'e}rez}}, \bibinfo {author}
  {\bibfnamefont {M.~C.}\ \bibnamefont {Cassidy}}, \bibinfo {author}
  {\bibfnamefont {S.}~\bibnamefont {Koelling}}, \bibinfo {author}
  {\bibfnamefont {S.}~\bibnamefont {Goswami}}, \bibinfo {author} {\bibfnamefont
  {K.}~\bibnamefont {Watanabe}}, \bibinfo {author} {\bibfnamefont
  {T.}~\bibnamefont {Taniguchi}}, \ and\ \bibinfo {author} {\bibfnamefont
  {L.~P.}\ \bibnamefont {Kouwenhoven}},\ }\href {\doibase 10.1038/ncomms16025}
  {\bibfield  {journal} {\bibinfo  {journal} {Nature Commun.}\ }\textbf
  {\bibinfo {volume} {8}},\ \bibinfo {pages} {16025} (\bibinfo {year}
  {2017})}\BibitemShut {NoStop}%
\bibitem [{\citenamefont {Gazibegovic}\ \emph {et~al.}(2017)\citenamefont
  {Gazibegovic}, \citenamefont {Car}, \citenamefont {Zhang}, \citenamefont
  {Balk}, \citenamefont {Logan}, \citenamefont {Moor}, \citenamefont {Cassidy},
  \citenamefont {Schmits}, \citenamefont {Xu}, \citenamefont {Wang},
  \citenamefont {Krogstrup}, \citenamefont {Veld}, \citenamefont {Zuo},
  \citenamefont {Vos}, \citenamefont {Shen}, \citenamefont {Bouman},
  \citenamefont {Shojaei}, \citenamefont {Pennachio}, \citenamefont {Lee},
  \citenamefont {Veldhoven}, \citenamefont {Koelling}, \citenamefont
  {Verheijen}, \citenamefont {Kouwenhoven}, \citenamefont {Palmstr{\o}m},\ and\
  \citenamefont {Bakkers}}]{gazibegovic_epitaxy_2017}%
  \BibitemOpen
  \bibfield  {author} {\bibinfo {author} {\bibfnamefont {S.}~\bibnamefont
  {Gazibegovic}}, \bibinfo {author} {\bibfnamefont {D.}~\bibnamefont {Car}},
  \bibinfo {author} {\bibfnamefont {H.}~\bibnamefont {Zhang}}, \bibinfo
  {author} {\bibfnamefont {S.~C.}\ \bibnamefont {Balk}}, \bibinfo {author}
  {\bibfnamefont {J.~A.}\ \bibnamefont {Logan}}, \bibinfo {author}
  {\bibfnamefont {M.~W. A.~d.}\ \bibnamefont {Moor}}, \bibinfo {author}
  {\bibfnamefont {M.~C.}\ \bibnamefont {Cassidy}}, \bibinfo {author}
  {\bibfnamefont {R.}~\bibnamefont {Schmits}}, \bibinfo {author} {\bibfnamefont
  {D.}~\bibnamefont {Xu}}, \bibinfo {author} {\bibfnamefont {G.}~\bibnamefont
  {Wang}}, \bibinfo {author} {\bibfnamefont {P.}~\bibnamefont {Krogstrup}},
  \bibinfo {author} {\bibfnamefont {R.~L. M. O.~h.}\ \bibnamefont {Veld}},
  \bibinfo {author} {\bibfnamefont {K.}~\bibnamefont {Zuo}}, \bibinfo {author}
  {\bibfnamefont {Y.}~\bibnamefont {Vos}}, \bibinfo {author} {\bibfnamefont
  {J.}~\bibnamefont {Shen}}, \bibinfo {author} {\bibfnamefont {D.}~\bibnamefont
  {Bouman}}, \bibinfo {author} {\bibfnamefont {B.}~\bibnamefont {Shojaei}},
  \bibinfo {author} {\bibfnamefont {D.}~\bibnamefont {Pennachio}}, \bibinfo
  {author} {\bibfnamefont {J.~S.}\ \bibnamefont {Lee}}, \bibinfo {author}
  {\bibfnamefont {P.~J.~v.}\ \bibnamefont {Veldhoven}}, \bibinfo {author}
  {\bibfnamefont {S.}~\bibnamefont {Koelling}}, \bibinfo {author}
  {\bibfnamefont {M.~A.}\ \bibnamefont {Verheijen}}, \bibinfo {author}
  {\bibfnamefont {L.~P.}\ \bibnamefont {Kouwenhoven}}, \bibinfo {author}
  {\bibfnamefont {C.~J.}\ \bibnamefont {Palmstr{\o}m}}, \ and\ \bibinfo
  {author} {\bibfnamefont {E.~P. A.~M.}\ \bibnamefont {Bakkers}},\ }\href
  {\doibase 10.1038/nature23468} {\bibfield  {journal} {\bibinfo  {journal}
  {Nature}\ }\textbf {\bibinfo {volume} {548}},\ \bibinfo {pages} {434}
  (\bibinfo {year} {2017})}\BibitemShut {NoStop}%
\bibitem [{\citenamefont {Aharonov}\ and\ \citenamefont
  {Bohm}(1959)}]{aharonov_significance_1959}%
  \BibitemOpen
  \bibfield  {author} {\bibinfo {author} {\bibfnamefont {Y.}~\bibnamefont
  {Aharonov}}\ and\ \bibinfo {author} {\bibfnamefont {D.}~\bibnamefont
  {Bohm}},\ }\href {\doibase 10.1103/PhysRev.115.485} {\bibfield  {journal}
  {\bibinfo  {journal} {Phys. Rev.}\ }\textbf {\bibinfo {volume} {115}},\
  \bibinfo {pages} {485} (\bibinfo {year} {1959})}\BibitemShut {NoStop}%
\bibitem [{\citenamefont {Webb}\ \emph {et~al.}(1985)\citenamefont {Webb},
  \citenamefont {Washburn}, \citenamefont {Umbach},\ and\ \citenamefont
  {Laibowitz}}]{webb_observation_1985}%
  \BibitemOpen
  \bibfield  {author} {\bibinfo {author} {\bibfnamefont {R.~A.}\ \bibnamefont
  {Webb}}, \bibinfo {author} {\bibfnamefont {S.}~\bibnamefont {Washburn}},
  \bibinfo {author} {\bibfnamefont {C.~P.}\ \bibnamefont {Umbach}}, \ and\
  \bibinfo {author} {\bibfnamefont {R.~B.}\ \bibnamefont {Laibowitz}},\ }\href
  {\doibase 10.1103/PhysRevLett.54.2696} {\bibfield  {journal} {\bibinfo
  {journal} {Phys. Rev. Lett.}\ }\textbf {\bibinfo {volume} {54}},\ \bibinfo
  {pages} {2696} (\bibinfo {year} {1985})}\BibitemShut {NoStop}%
\bibitem [{\citenamefont {Timp}\ \emph {et~al.}(1987)\citenamefont {Timp},
  \citenamefont {Chang}, \citenamefont {Cunningham}, \citenamefont {Chang},
  \citenamefont {Mankiewich}, \citenamefont {Behringer},\ and\ \citenamefont
  {Howard}}]{timp_observation_1987}%
  \BibitemOpen
  \bibfield  {author} {\bibinfo {author} {\bibfnamefont {G.}~\bibnamefont
  {Timp}}, \bibinfo {author} {\bibfnamefont {A.~M.}\ \bibnamefont {Chang}},
  \bibinfo {author} {\bibfnamefont {J.~E.}\ \bibnamefont {Cunningham}},
  \bibinfo {author} {\bibfnamefont {T.~Y.}\ \bibnamefont {Chang}}, \bibinfo
  {author} {\bibfnamefont {P.}~\bibnamefont {Mankiewich}}, \bibinfo {author}
  {\bibfnamefont {R.}~\bibnamefont {Behringer}}, \ and\ \bibinfo {author}
  {\bibfnamefont {R.~E.}\ \bibnamefont {Howard}},\ }\href {\doibase
  10.1103/PhysRevLett.58.2814} {\bibfield  {journal} {\bibinfo  {journal}
  {Phys. Rev. Lett.}\ }\textbf {\bibinfo {volume} {58}},\ \bibinfo {pages}
  {2814} (\bibinfo {year} {1987})}\BibitemShut {NoStop}%
\bibitem [{\citenamefont {Groth}\ \emph {et~al.}(2014)\citenamefont {Groth},
  \citenamefont {Wimmer}, \citenamefont {Akhmerov},\ and\ \citenamefont
  {Waintal}}]{groth_kwant:_2014}%
  \BibitemOpen
  \bibfield  {author} {\bibinfo {author} {\bibfnamefont {C.~W.}\ \bibnamefont
  {Groth}}, \bibinfo {author} {\bibfnamefont {M.}~\bibnamefont {Wimmer}},
  \bibinfo {author} {\bibfnamefont {A.~R.}\ \bibnamefont {Akhmerov}}, \ and\
  \bibinfo {author} {\bibfnamefont {X.}~\bibnamefont {Waintal}},\ }\href
  {\doibase 10.1088/1367-2630/16/6/063065} {\bibfield  {journal} {\bibinfo
  {journal} {New J. Phys.}\ }\textbf {\bibinfo {volume} {16}},\ \bibinfo
  {pages} {063065} (\bibinfo {year} {2014})}\BibitemShut {NoStop}%
\bibitem [{\citenamefont {Beenakker}(1992)}]{beenakker_quantum_1992}%
  \BibitemOpen
  \bibfield  {author} {\bibinfo {author} {\bibfnamefont {C.~W.~J.}\
  \bibnamefont {Beenakker}},\ }\href {\doibase 10.1103/PhysRevB.46.12841}
  {\bibfield  {journal} {\bibinfo  {journal} {Phys. Rev. B}\ }\textbf {\bibinfo
  {volume} {46}},\ \bibinfo {pages} {12841} (\bibinfo {year}
  {1992})}\BibitemShut {NoStop}%
\bibitem [{\citenamefont {van Weperen}\ \emph {et~al.}(2015)\citenamefont {van
  Weperen}, \citenamefont {Tarasinski}, \citenamefont {Eeltink}, \citenamefont
  {Pribiag}, \citenamefont {Plissard}, \citenamefont {Bakkers}, \citenamefont
  {Kouwenhoven},\ and\ \citenamefont {Wimmer}}]{van_weperen_spin-orbit_2015}%
  \BibitemOpen
  \bibfield  {author} {\bibinfo {author} {\bibfnamefont {I.}~\bibnamefont {van
  Weperen}}, \bibinfo {author} {\bibfnamefont {B.}~\bibnamefont {Tarasinski}},
  \bibinfo {author} {\bibfnamefont {D.}~\bibnamefont {Eeltink}}, \bibinfo
  {author} {\bibfnamefont {V.~S.}\ \bibnamefont {Pribiag}}, \bibinfo {author}
  {\bibfnamefont {S.~R.}\ \bibnamefont {Plissard}}, \bibinfo {author}
  {\bibfnamefont {E.~P. A.~M.}\ \bibnamefont {Bakkers}}, \bibinfo {author}
  {\bibfnamefont {L.~P.}\ \bibnamefont {Kouwenhoven}}, \ and\ \bibinfo {author}
  {\bibfnamefont {M.}~\bibnamefont {Wimmer}},\ }\href {\doibase
  10.1103/PhysRevB.91.201413} {\bibfield  {journal} {\bibinfo  {journal} {Phys.
  Rev. B}\ }\textbf {\bibinfo {volume} {91}},\ \bibinfo {pages} {201413}
  (\bibinfo {year} {2015})}\BibitemShut {NoStop}%
\bibitem [{\citenamefont {Kammhuber}\ \emph {et~al.}(2017)\citenamefont
  {Kammhuber}, \citenamefont {Cassidy}, \citenamefont {Pei}, \citenamefont
  {Nowak}, \citenamefont {Vuik}, \citenamefont {G{\"u}l}, \citenamefont {Car},
  \citenamefont {Plissard}, \citenamefont {Bakkers}, \citenamefont {Wimmer},\
  and\ \citenamefont {Kouwenhoven}}]{kammhuber_conductance_2017}%
  \BibitemOpen
  \bibfield  {author} {\bibinfo {author} {\bibfnamefont {J.}~\bibnamefont
  {Kammhuber}}, \bibinfo {author} {\bibfnamefont {M.~C.}\ \bibnamefont
  {Cassidy}}, \bibinfo {author} {\bibfnamefont {F.}~\bibnamefont {Pei}},
  \bibinfo {author} {\bibfnamefont {M.~P.}\ \bibnamefont {Nowak}}, \bibinfo
  {author} {\bibfnamefont {A.}~\bibnamefont {Vuik}}, \bibinfo {author}
  {\bibfnamefont {{\"O}.}~\bibnamefont {G{\"u}l}}, \bibinfo {author}
  {\bibfnamefont {D.}~\bibnamefont {Car}}, \bibinfo {author} {\bibfnamefont
  {S.~R.}\ \bibnamefont {Plissard}}, \bibinfo {author} {\bibfnamefont {E.~P.
  a.~M.}\ \bibnamefont {Bakkers}}, \bibinfo {author} {\bibfnamefont
  {M.}~\bibnamefont {Wimmer}}, \ and\ \bibinfo {author} {\bibfnamefont {L.~P.}\
  \bibnamefont {Kouwenhoven}},\ }\href {\doibase 10.1038/s41467-017-00315-y}
  {\bibfield  {journal} {\bibinfo  {journal} {Nature Commun.}\ }\textbf
  {\bibinfo {volume} {8}},\ \bibinfo {pages} {478} (\bibinfo {year}
  {2017})}\BibitemShut {NoStop}%
\bibitem [{\citenamefont {W{\'o}jcik}, \citenamefont {Bertoni},\ and\
  \citenamefont {Goldoni}(2018{\natexlab{a}})}]{wojcik_tuning_2018}%
  \BibitemOpen
  \bibfield  {author} {\bibinfo {author} {\bibfnamefont {P.}~\bibnamefont
  {W{\'o}jcik}}, \bibinfo {author} {\bibfnamefont {A.}~\bibnamefont {Bertoni}},
  \ and\ \bibinfo {author} {\bibfnamefont {G.}~\bibnamefont {Goldoni}},\ }\href
  {\doibase 10.1103/PhysRevB.97.165401} {\bibfield  {journal} {\bibinfo
  {journal} {Phys. Rev. B}\ }\textbf {\bibinfo {volume} {97}},\ \bibinfo
  {pages} {165401} (\bibinfo {year} {2018}{\natexlab{a}})}\BibitemShut
  {NoStop}%
\bibitem [{\citenamefont {Sticlet}, \citenamefont {Nijholt},\ and\
  \citenamefont {Akhmerov}(2017)}]{sticlet_robustness_2017}%
  \BibitemOpen
  \bibfield  {author} {\bibinfo {author} {\bibfnamefont {D.}~\bibnamefont
  {Sticlet}}, \bibinfo {author} {\bibfnamefont {B.}~\bibnamefont {Nijholt}}, \
  and\ \bibinfo {author} {\bibfnamefont {A.}~\bibnamefont {Akhmerov}},\ }\href
  {\doibase 10.1103/PhysRevB.95.115421} {\bibfield  {journal} {\bibinfo
  {journal} {Phys. Rev. B}\ }\textbf {\bibinfo {volume} {95}},\ \bibinfo
  {pages} {115421} (\bibinfo {year} {2017})}\BibitemShut {NoStop}%
\bibitem [{\citenamefont {Nowak}\ and\ \citenamefont
  {W{\'o}jcik}(2018)}]{nowak_renormalization_2018}%
  \BibitemOpen
  \bibfield  {author} {\bibinfo {author} {\bibfnamefont {M.~P.}\ \bibnamefont
  {Nowak}}\ and\ \bibinfo {author} {\bibfnamefont {P.}~\bibnamefont
  {W{\'o}jcik}},\ }\href {\doibase 10.1103/PhysRevB.97.045419} {\bibfield
  {journal} {\bibinfo  {journal} {Phys. Rev. B}\ }\textbf {\bibinfo {volume}
  {97}},\ \bibinfo {pages} {045419} (\bibinfo {year} {2018})}\BibitemShut
  {NoStop}%
\bibitem [{Note1()}]{Note1}%
  \BibitemOpen
  \bibinfo {note} {The choice of a vector potential that minimizes the
  supercurrent in the considered device is vital for proper description of the
  orbital effects of the magnetic field, see Ref. \protect \rev@citealpnum
  {wojcik_durability_2018}}\BibitemShut {NoStop}%
\bibitem [{\citenamefont {Antipov}\ \emph {et~al.}(2018)\citenamefont
  {Antipov}, \citenamefont {Bargerbos}, \citenamefont {Winkler}, \citenamefont
  {Bauer}, \citenamefont {Rossi},\ and\ \citenamefont
  {Lutchyn}}]{antipov_effects_2018}%
  \BibitemOpen
  \bibfield  {author} {\bibinfo {author} {\bibfnamefont {A.~E.}\ \bibnamefont
  {Antipov}}, \bibinfo {author} {\bibfnamefont {A.}~\bibnamefont {Bargerbos}},
  \bibinfo {author} {\bibfnamefont {G.~W.}\ \bibnamefont {Winkler}}, \bibinfo
  {author} {\bibfnamefont {B.}~\bibnamefont {Bauer}}, \bibinfo {author}
  {\bibfnamefont {E.}~\bibnamefont {Rossi}}, \ and\ \bibinfo {author}
  {\bibfnamefont {R.~M.}\ \bibnamefont {Lutchyn}},\ }\href {\doibase
  10.1103/PhysRevX.8.031041} {\bibfield  {journal} {\bibinfo  {journal} {Phys.
  Rev. X}\ }\textbf {\bibinfo {volume} {8}},\ \bibinfo {pages} {031041}
  (\bibinfo {year} {2018})}\BibitemShut {NoStop}%
\bibitem [{\citenamefont {Winkler}\ \emph {et~al.}(2018)\citenamefont
  {Winkler}, \citenamefont {Antipov}, \citenamefont {van Heck}, \citenamefont
  {Soluyanov}, \citenamefont {Glazman}, \citenamefont {Wimmer},\ and\
  \citenamefont {Lutchyn}}]{winkler_unified_2018}%
  \BibitemOpen
  \bibfield  {author} {\bibinfo {author} {\bibfnamefont {G.~W.}\ \bibnamefont
  {Winkler}}, \bibinfo {author} {\bibfnamefont {A.~E.}\ \bibnamefont
  {Antipov}}, \bibinfo {author} {\bibfnamefont {B.}~\bibnamefont {van Heck}},
  \bibinfo {author} {\bibfnamefont {A.~A.}\ \bibnamefont {Soluyanov}}, \bibinfo
  {author} {\bibfnamefont {L.~I.}\ \bibnamefont {Glazman}}, \bibinfo {author}
  {\bibfnamefont {M.}~\bibnamefont {Wimmer}}, \ and\ \bibinfo {author}
  {\bibfnamefont {R.~M.}\ \bibnamefont {Lutchyn}},\ }\href
  {http://arxiv.org/abs/1810.04180} {\bibfield  {journal} {\bibinfo  {journal}
  {arXiv:1810.04180}\ } (\bibinfo {year} {2018})}\BibitemShut {NoStop}%
\bibitem [{\citenamefont {W{\'o}jcik}, \citenamefont {Bertoni},\ and\
  \citenamefont {Goldoni}(2018{\natexlab{b}})}]{wojcik_enhanced_2018}%
  \BibitemOpen
  \bibfield  {author} {\bibinfo {author} {\bibfnamefont {P.}~\bibnamefont
  {W{\'o}jcik}}, \bibinfo {author} {\bibfnamefont {A.}~\bibnamefont {Bertoni}},
  \ and\ \bibinfo {author} {\bibfnamefont {G.}~\bibnamefont {Goldoni}},\ }\href
  {http://arxiv.org/abs/1811.09088} {\bibfield  {journal} {\bibinfo  {journal}
  {arXiv:1811.09088}\ } (\bibinfo {year} {2018}{\natexlab{b}})}\BibitemShut
  {NoStop}%
\bibitem [{\citenamefont {W{\'o}jcik}\ and\ \citenamefont
  {Nowak}(2018)}]{wojcik_durability_2018}%
  \BibitemOpen
  \bibfield  {author} {\bibinfo {author} {\bibfnamefont {P.}~\bibnamefont
  {W{\'o}jcik}}\ and\ \bibinfo {author} {\bibfnamefont {M.~P.}\ \bibnamefont
  {Nowak}},\ }\href {\doibase 10.1103/PhysRevB.97.235445} {\bibfield  {journal}
  {\bibinfo  {journal} {Phys. Rev. B}\ }\textbf {\bibinfo {volume} {97}},\
  \bibinfo {pages} {235445} (\bibinfo {year} {2018})}\BibitemShut {NoStop}%
\end{thebibliography}%
\end{document}